# Near-infrared scanning cavity ringdown for optical loss characterization of supermirrors

G. W. Truong[1]*, G. Winkler[2], T. Zederbauer[3], D. Bachmann[3], P. Heu[1], D. Follman[1], M. E. White[1], O. H. Heckl[2], G. D. Cole[1,3]

[1] *Crystalline Mirror Solutions LLC, Santa Barbara, California 93101, USA*
[2] *Christian Doppler Laboratory for Mid-IR Spectroscopy and Semiconductor Optics, Faculty Center for Nano Structure Research, Faculty of Physics, University of Vienna, 1090 Vienna, Austria*
[3] *Crystalline Mirror Solutions GmbH, 1010 Vienna, Austria*
* *garwing@crystallinemirrors.com*

**Abstract:** A cavity ringdown system for probing the spatial variation of optical loss across high-reflectivity mirrors is described. This system is employed to examine substrate-transferred crystalline supermirrors and quantify the effect of manufacturing process imperfections. Excellent agreement is observed between the ringdown-generated spatial measurements and differential interference contrast microscopy images. A 2 mm diameter ringdown scan in the center of a crystalline supermirror reveals highly uniform coating properties with excess loss variations below 1 ppm.

## 1. Introduction

Substrate-transferred crystalline coatings are a new class of optical interference coatings shown to have equivalent optical properties to high-performance amorphous coatings in the near-infrared, but with a tenfold reduction of Brownian noise [1] at room temperature when compared to ion-beam sputtered (IBS) multilayers based on $SiO_2$ and $Ta_2O_5$ films [2]. Consequently, a variety of thermal-noise-limited applications of precision optical interferometry benefit from crystalline coatings. In particular, state-of-the-art designs for ultrastable reference cavities are now limited by the thermal noise of the mirror coatings [3,4], which can be improved by using low-elastic loss materials such as monocrystalline GaAs/AlGaAs Bragg reflectors [1]. Similarly, the performance of large-area ring laser gyroscopes [5,6] and gravitational wave detectors [7,8] can potentially be improved by the adoption of coatings with less Brownian noise contribution.

Recognizing that the substrate-transfer fabrication technique differs greatly from conventional coating methods such as physical vapor deposition, it is useful to characterize the optical losses and study their uniformity across the entirety of the coating. In an initial study, a small sample set of cavity ringdown data taken at multiple locations on a pair of crystalline coatings revealed a small variation of optical loss at the ~10 ppm level at isolated points [9]. In a more recent study performed on 50.8-mm (2-inch) diameter crystalline optics, optical inspection showed that the surface density for defects smaller than 100 μm was similar to those of the IBS coatings currently used at the LIGO and Virgo gravitational wave detectors [10]. However, non-ideal surface properties of both the epitaxial material and the optical substrate caused a higher density of defects sized greater than 100 μm [10]. The inspection techniques employed in [10] covered large coating areas in a short time but inherently lack the link between the analyzed optical inspection data and excess optical losses.

In this paper, we describe a method to systematically and directly probe the optical loss using an automated scanning cavity ringdown technique. The optical setup, consisting of free-running and unisolated laser diodes directly coupled into linear cavities, provides measurement flexibility for multiple wavelengths (thus far we have employed 1064 nm, 1156 nm, 1397 nm,

1550 nm, and 1572 nm). Our approach is complementary to that of Cui et. al. [11] and Han et. al. [12], who used a non-linear cavity configuration. Tan et. al. [13] reconstructed the spatial loss distribution map based on the losses experienced by multiple orthogonal higher-order cavity modes. The advantage of this non-linear cavity configuration is its ability to spatially separate the cavity-coupled and rejected modes such that only light that is resonant with the cavity is fed back to the laser [14,15]. Cui et. al. and Han et. al. suggest that this mode stability contributed greatly to the tight statistics seen in their measurements.

In the present case, a linear cavity is used to reduce the amount of mounting hardware and thus the volume of the vacuum envelope required, potentially at the detriment to mode stability. In contrast to the approach by Tan et. al. in which the loss distribution from each of the two mirrors cannot be easily separated, we are able to map the losses over a single mirror by scanning one mirror and leaving the other one at a fixed position. For crystalline coatings, exquisite layer thickness control and post-growth characterization abilities in the semiconductor toolset allow the coating transmission to be determined with ~1 ppm uncertainty. In essence, this precision turns the ringdown measurements into a direct probe of excess optical loss, i.e., scatter plus absorption. Knowledge of the minimum-achievable excess loss is the parameter of interest for many experiments, because this will place a lower bound for the design value of transmission to achieve the highest practically-usable cavity finesse. Further partitioning of the excess loss into its constituent components can be performed using photothermal common-path interferometry (PCI) to measure absorption loss [16,17].

To our knowledge, this letter is also the first demonstration of automated scanning of ultra-low-loss planar and curved mirrors with total optical losses below 20 ppm (i.e., reflectivity >99.998%). In addition, the ringdown results are compared to micrographs generated via differential interference contrast (DIC) microscopy to identify the physical cause of the spatial inhomogeneity of the measured optical loss. The ability to quantify the spatial homogeneity of low-loss coatings is critical to many high-precision applications such as the construction of optical reference cavities, qualifying large optics in gravitational wave detectors, or for cavity ringdown spectroscopy.

## 2. Scanning cavity ringdown

Figure 1(a) shows a schematic of the scanning cavity ringdown setup. For rapid and wide wavelength interchangeability in the near-infrared, we use readily-available Fabry-Perot (FP) and distributed feedback (DFB) diode lasers in a universal temperature- and current-controlled mount. The diodes are used in an unisolated configuration without active frequency stabilization. The cavity input coupler, typically a planar mirror, is aligned for maximal optical feedback, thereby forming an extended cavity diode laser (ECDL). This arrangement narrows the laser linewidth, which increases the in-coupled optical power, and pulls the laser's center wavelength to one of the modes of the passive cavity as seen in Figure 1(a). The modes of the passive cavity will be determined by the coating center wavelength and bandwidth. The ultimate lasing wavelength will then be at the part of the spectrum where the extended cavity laser has least loss consistent with the bare laser gain bandwidth and external mirrors. We have chosen to use this optical-feedback-assisted cavity ringdown technique for the combined benefits of experimental simplicity and associated cost reduction while maintaining high transmitted power for good measurement precision.

Two gimballed mirror holders on motorized linear stages all with encoded actuators are mounted inside a vacuum chamber (Figure 1(b)). The vacuum chamber, which can reach an ultimate pressure of ~$10^{-5}$ torr without special preparation, enables the characterization of the mirror losses at wavelengths at which there is significant atmospheric absorption. Different locations on the coating can be probed by moving the two-axis translation stage. For mapping over non-planar substrates of radius-of-curvature $R$, an additional adjustment of the tip or tilt angle of the mirror is required after translation. Supposing that the mirror normal is initially colinear with the beam axis through the cavity, a translation of distance $\rho$ in the plane orthogonal to the beam will result in the beam now sampling a new part of the mirror with a normal that is deviated by $\theta \approx \rho/R$. The linear actuators have a typical positional accuracy of 2.2 μm, which is converted to an angular accuracy of $1.4 \times 10^{-3}$ degrees (via the gimbal mount positioning conversion of $G \approx 1.55$ mm/deg). For comparison, typical values of $\rho = 0.1$ mm and $R = 1000$ mm requires an angular compensation of $5.7 \times 10^{-3}$ degrees. In this initial demonstration, we have chosen to use a feed-forward strategy for curvature compensation with no further optimization.

Most of the cavity-transmitted beam is directed to a fast InGaAs photodiode whose fall time is much shorter than 2.5 μs, allowing cavities of length 10 cm with finesse as low as 24,000 to be measured. Input alignment coupling is optimized for the fundamental cavity mode, but higher-order modes are often still excited with a transmission typically below one-tenth of the fundamental. Care should be taken so that finesse measurements are not made from modes with vastly different spatial intensity distribution that differently sample mirror defects. However, since the laser and cavity are not actively locked to each other, resonance occurs in a non-deterministic fashion. Given this random cavity excitation, we constructed a custom digital delay generator [18] to modulate the laser diode current to zero when the transmitted power exceeded a threshold voltage, and to trigger data acquisition of a single ringdown transient. The threshold level is chosen to exclude events from high-order modes, but image analysis of the transmitted mode is also performed for redundancy. Figure 1(e) shows an example of a single ringdown captured in this way and the residuals generated from a least squares fit to the model $y = a\, e^{-t/\tau} + b$, where $t$ is laboratory time, $\{a,b,\tau\}$ are free parameters, and $\tau$ is the optical decay time. Additionally, the average of 50 consecutive ringdowns and their fit residuals are displayed, and there are no signs of non-exponential behavior (which can arise from mode beating and cross-coupling due to scattering) at our highest levels of signal-to-noise ratio. Using a value of 1.39 mV estimated from the standard deviation of the voltage noise, we computed the reduced chi-squared statistic to be $\chi_v^2 = 1.00$, indicating an excellent fit to a single exponential. The statistical error in the fitted value of $\tau$ was 0.2% computed from the goodness of fit. The optical decay time is converted to an optical loss knowing the mirror separation $L$, and assuming equal loss on each mirror using the relation $T + S + A = \pi/F$, where $F = c\pi\tau/L$ is the finesse, $T$ is the transmission, $S$ is the scatter loss, $A$ is the absorption loss, each defined per mirror, and $c$ is the speed of light. Typically, $L = 92$ mm and the spot size on a planar and 1-m radius-of-curvature mirror are $w = 313$ μm and 328 μm, respectively, for a cavity at 1064 nm. Since the linear actuators provide much finer positional accuracy, we will consider the spatial resolution of this apparatus to be limited by the laser spot size.

About 10% of the beam power transmitted through the cavity is split onto an InGaAs camera. Image processing is used to determine whether the observed mode is the fundamental TEM$_{00}$ mode. After determining the contour of the optical mode, we use the circularity metric, $C = 4\pi A/l^2$, which has a maximum value of 1 if and only if the contour is a circle. $A$ and $l$ are the enclosed area and perimeter of the mode contour, respectively. The circularity metric is invariant to the rather significant instability in transmitted power through the unstabilized cavity. Allowing for some non-uniform image background caused by scattered and stray light, we typically consider values of $C > 0.85$ as a true fundamental mode. We note that more robust algorithms can be implemented if required. For example, least-squares fitting to a family of cavity spatial eigenmodes could potentially provide better mode determination. To provide some immunity against unstable mode coupling (due to an unstabilized laser and cavity length), we typically average 100 frames (over a few seconds) which provides a probability $P = N_{\text{fund}}/N_{\text{frames}}$ that the current coupling configuration excites the fundamental mode. $N_{\text{fund}}$ is the number of frames in which $C > 0.85$ and $N_{\text{frames}}$ is the total number of non-trivial frames (i.e., with $A > 0$). Figure 1(c-d) shows examples of frames analyzed in this way. The values of $P$ taken at each point on the mirrors are recorded and can be used in future re-optimization algorithms to re-align the tip/tilt of curved mirrors after translation to re-find the fundamental mode.

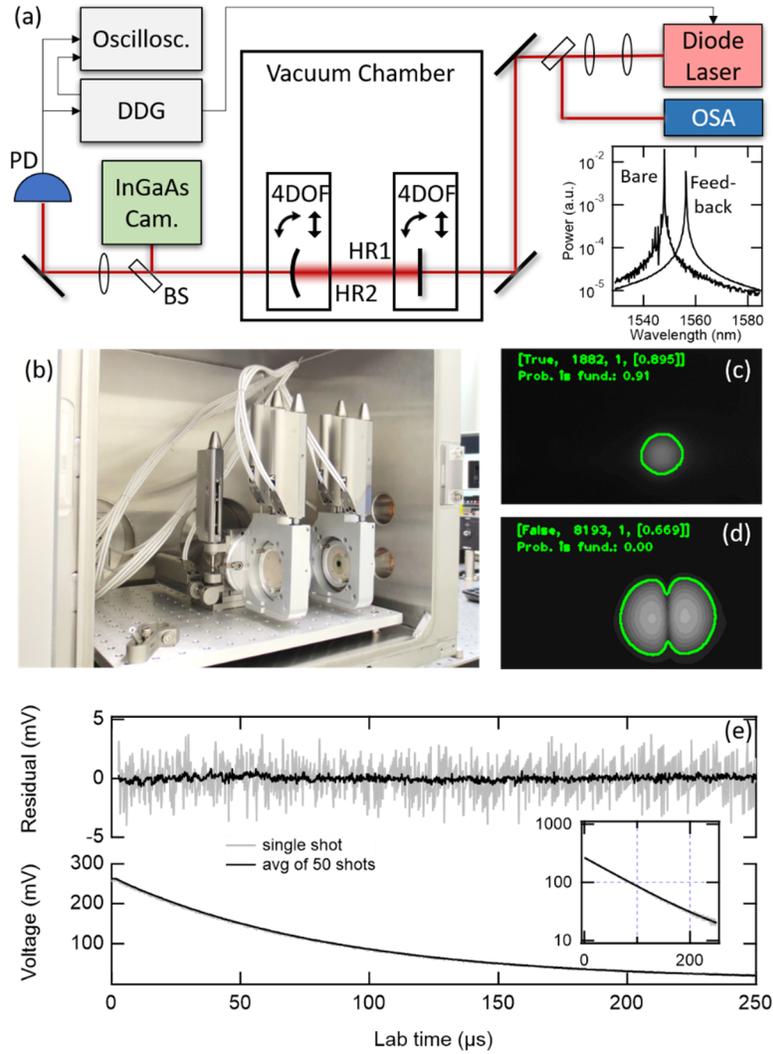

Figure 1. (a) Schematic of the scanning ringdown apparatus. The inset shows the shift of the center wavelength of the diode under bare-lasing and passive feedback conditions. HR1, HR2: High-reflectivity mirrors under test; 4DOF: Four degree-of-freedom stages allowing transverse translation and tip/tilt adjustment; BS: beamsplitter; PD: photodetector; OSA: optical spectrum analyzer; DDG: digital delay generator. (b) Photograph showing the 4DOF stages and motorized actuators inside the vacuum chamber. (c-d) Examples of frames captured by the InGaAs camera showing the fundamental mode with $C = 0.895$ and $P = 0.91$, and the TEM01 mode with $C = 0.669$ and $P = 0.00$. (e) Time domain data showing a single ringdown instance and the fit residuals (grey). The reduced chi-squared statistic is $\chi_\nu^2 = 1.00$. The average of 50 ringdowns and residuals are also shown (black). The inset shows the ringdown signal on a log-linear scale.

We established the reproducibility of the loss measurement at a fixed location by taking repeated ringdowns without moving either mirror. In Figure 2(a), we show the normalized histograms from a variety of realizations of the experiment with the different diodes (which have large variations in the output beam quality that affects spatial mode stability). All sources were Fabry-Perot-type diodes with the exception of the 1156 nm distributed feedback diode laser. The typical standard deviation of the loss measurements was 0.1 ppm or less, with the exception being the 1064 nm measurements in which the spread was 1 ppm. For the sample size of 100 repeated measurements, the standard error of the mean provides a precision of 0.01 ppm, but drifts begin to become apparent at this level. We hypothesize that the difference in statistics is driven by the spatial mode stability of the 1064 nm diode laser under optical feedback since this is known to be a strong contributor to the measurement noise [11,12]. In Figure 2(b), we explored the spatial reproducibility by repeating measurements over a particular high-loss feature. For 100 μm-spaced positional sampling, there was no appreciable spatial reproducibility error seen with a probe beam size of ~ 300 μm.

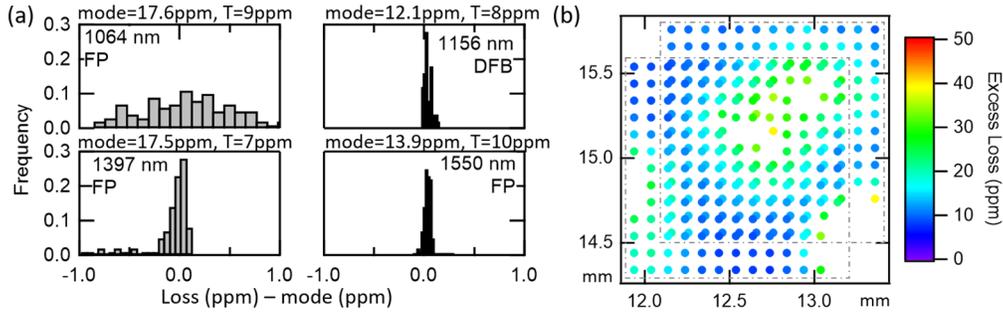

Figure 2. (a) Histograms showing that the repeatability of the loss measurement in a static experiment is somewhat dependent on the laser diode source. The sample size in each case is 100. The sample mode (i.e., the most frequently-appearing value of the distribution) is subtracted from each distribution to show the difference in spread for the loss measurement in each case. The values of the mode and assumed transmission are shown in ppm above each plot. (b) Repeated scans (indicated by the grey dashed boxes) covering a defective spot showed excellent spatial reproducibility. One dataset has been intentionally offset in the horizontal and vertical axes by 0.02 mm for clarity. $T$ = 9 ppm.

3. **Uniformity measurements**

To demonstrate the performance of our mapping system, we measured the losses across a rejected 8-mm diameter crystalline coating transferred to a planar fused silica substrate. This sample was selected for this experiment because of its easily identifiable defect patterns. A 1-m radius-of-curvature (ROC-1) mirror with an identical HR coating was used as a reference mirror to form a 92-mm-long cavity. The curved mirror remained stationary throughout the experiment with the beam sampling a defect-free area near its center. Figure 3(a) shows the measured excess optical losses overlaid on a stitched DIC image highlighting the locations of visible defects. X-ray diffraction measurements of the GaAs/AlGaAs coating enables the as-grown layer thicknesses to be determined and the known refractive indices for the coating and substrate materials permit accurate calculation of the transmission (9 ppm) at the design wavelength. This transmission was subtracted from the total measured loss. To avoid strong effects from outlying measurements caused by infrequent large perturbations of the unstabilized laser, the median loss from a set of ten ringdowns is attributed to each point in Figure 3(a). Repeated measurements over a period of one week showed that the results are robust against drifts and imperfections in the mechanical systems.

A magnified view of a region near the center of the coating is shown in Figure 3(b). We observed that there are two types of coating defects arising from two different physical origins unique to the substrate-transfer process. Epitaxial growth defects (known as "oval defects" arising from spitting of metals in the epitaxial deposition chamber [19]) lead to strong disruptions of the optical field resulting in a region of radius ~1.5$w$ where no resonant mode is supported. This result is in fair agreement with a calculated beam radius of 0.75 mm on the planar mirror that encloses 99.998% (i.e., 20 ppm excluded) of the beam power. These types of defects can readily be seen as large data-excluded regions in Figure 3(a). The second type of defect visible in Figure 3(b) is an extended void formed where the coating had not made complete contact with the substrate. These are less deleterious to the optical field, and their contribution to loss seems to be at a smaller magnitude and more spatially restricted. Nevertheless, we see that minor bond defects that are barely visible in DIC (Figure 3(c)) can lead to an additional ~3 ppm of loss.

Figure 3(d) shows that there was no global correlation between the spatial mode of the probe (quantified by the probability $P$) and the measured loss, indicating that no filtering or deconvolution of the map shown in Figure 3(a) to account for different mode sizes is necessary. In general, this would be true if, like in this dataset, few non-TEM$_{00}$ modes (with relatively low mode numbers) were excited with enough transmission to trigger ringdowns. In addition, the histogram shows that the most probable value of excess loss was 10 ppm. Independent PCI measurements on nominally identical coating material produced in the same semiconductor growth showed that about 70% of this excess loss was dominated by absorption. We note that absorption is largely a function of impurity dopant concentration and is reduced by more careful epitaxial growth conditions. We have measured a minimum of 0.6 ppm of absorption loss in other coating runs [9].

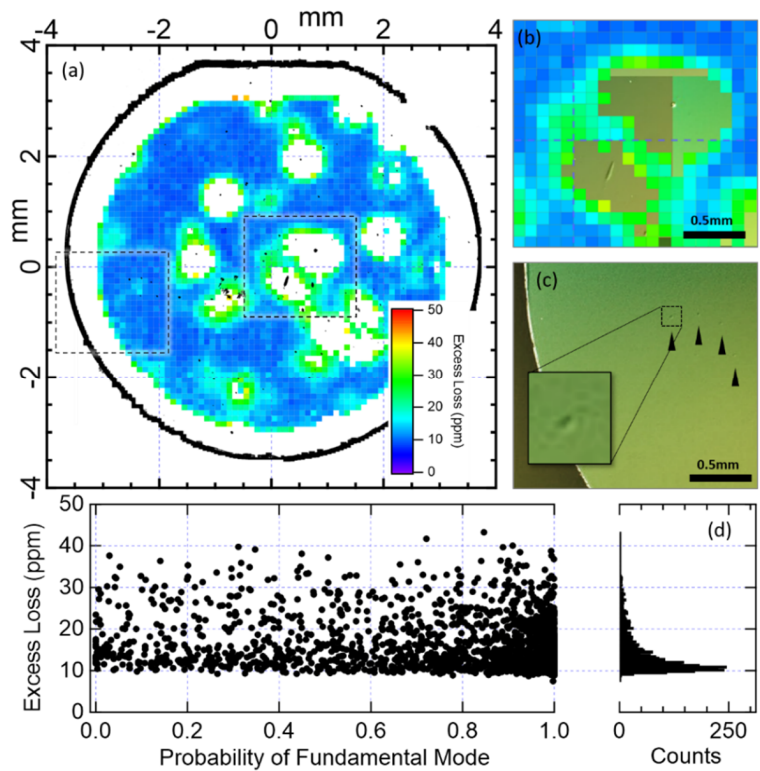

Figure 3. (a) Overlay of 8 mm diameter cavity ringdown measurements of optical loss taken with 0.1 mm point spacing (colored squares) on top of locations of defects inferred from DIC images (black). Locations where no ringdown data was available are transparent. The edge of the coating is treated as a "defect." $T$ = 9 ppm. (b) Zoom of ringdown measurements near the center of the coating on top of stitched DIC images. (c) DIC image of a region near the left edge of the coating with a series of small defects indicated by the arrows. (d) Scatter plot and histogram of all measured values of loss.

Figure 4 shows the measured loss over a quarter of the 8-mm-diameter coating that was transferred to a 1-m-radius-of-curvature substrate. The results are similar to that for an equivalent planar mirror (Figure 3(a)), except for some additional isolated dropouts in the data caused by an insufficient amount of light coupled into the cavity for measurement. We expect that a more refined determination of the value of $G$ and with a feedback strategy using the spatial mode information collected by the InGaAs camera will increase coverage and reduce dropouts at the cost of total mapping time.

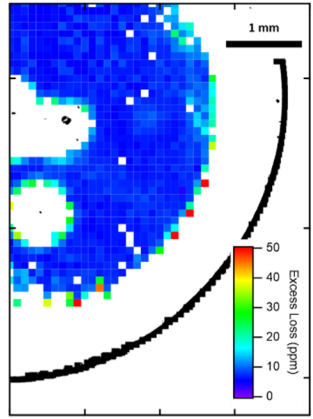

Figure 4. Overlay of optical loss taken with 0.1 mm point spacing (colored squares) on top of locations of defects inferred from DIC images (black) for a crystalline coating transferred to a substrate with a 1-m ROC. $T$ = 9 ppm

The same scanning cavity ringdown method was used to scan a region near the center of a production sample that showed fewer defects under the DIC microscope. Figure 5 shows that a 2 mm diameter clear aperture achieved minimal variation of losses in that region. We particularly use this mapping capability when producing mirrors intended for assembly into an optical cavity (which will typically have spot sizes < 1 mm at the mirror). In these cases, it is important to ensure that high finesse can be guaranteed under small variations in assembly tolerances (typically < 200 μm) and that the clear aperture coincides with the center of the substrate.

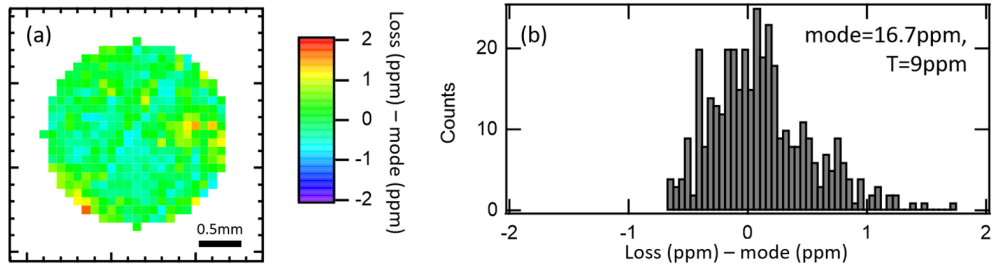

Figure 5. Production grade crystalline mirror at a wavelength of 1064 nm. The sample mode (i.e. the most frequently-appearing value of the distribution) is subtracted to show the variation of losses. (a) Example of a highly uniform coating with a clear aperture diameter of 2 mm. (b) Histogram of losses in the clear aperture, showing < 1 ppm (FWHM) variation across the surface.

## 4. Conclusions and outlook

Crystalline coatings have emerged as a completely new method for producing low-loss supermirrors suitable for use in a variety of interferometric and sensing applications. Continual process development has enabled a reduction of the excess optical loss from 17 ppm (reported in 2013) to 2.6 ppm in a recent coating run with a peak reflectivity near 1550 nm. However, a key requirement for the practical usability of any low-loss mirror is maintaining the low losses with high uniformity over the entire coating surface. Here, we have demonstrated the mapping of the optical losses across the entire area of a mirror and correlated the extracted values with micrographs obtained using DIC microscopy. With this system, a high-quality coating run was shown to have a variation in excess loss below 1 ppm. More broadly, the ability to measure the spatial variations in optical loss below the 20 ppm level on both planar and curved surfaces is critically important for characterizing supermirrors irrespective of fabrication technique.

**Funding and acknowledgements**

GW, TZ, DB, OHH gratefully acknowledge support by the Austrian Research Promotion Agency (FFG) within the projects 849731,861353, and 865556. GW and OHH gratefully acknowledge the financial support by the Austrian Federal Ministry for Digital and Economic Affairs and the National Foundation for Research, Technology and Development. The authors acknowledge the contributions of Blake Haist towards the development of the digital delay generator.